\documentclass[review]{elsarticle}
\usepackage{graphicx}
\usepackage{multirow}
\journal{Journal of Computational Physics}

\begin{document}

\begin{frontmatter}

\title{CROFT: A scalable three-dimensional parallel Fast Fourier Transform (FFT) implementation for High Performance Clusters}
\author{Vivek Gavane\fnref{myfootnote}\corref{mycorrespondingauthor}}
\ead{vivekg@cdac.in}
\author{Supriya Prabhugawankar\fnref{myfootnote}} 
\author{Shivam Garg} 
\author{Archana Achalere}
\author{Rajendra Joshi\corref{mycorrespondingauthor}}
\address{HPC-Medical and Bioinformatics Applications Group, Centre for Development of Advanced Computing(C-DAC), C-DAC Innovation Park, Pashan, Pune-411008, India.}
\cortext[mycorrespondingauthor]{Corresponding author}
\ead{rajendra@cdac.in}
\fntext[myfootnote]{Vivek Gavane and Supriya Prabhugawankar should be regarded as joint first authors}
\begin{abstract}
The FFT of three-dimensional (3D) input data is an important computational kernel of numerical simulations and is widely used in 
High Performance Computing (HPC) codes running on a large number of processors. Performance of many scientific applications such as 
Molecular Dynamic simulations depends on the underlying 3D parallel FFT library being used.  

In this paper, we present C-DAC’s three-dimensional Fast Fourier Transform (CROFT) library which implements three-dimensional parallel 
FFT using pencil decomposition. To exploit the hyperthreading capabilities of processor cores  without affecting performance, CROFT is 
designed to use multithreading along with MPI. CROFT implementation has an innovative feature of overlapping compute and memory-I/O 
with MPI communication using multithreading. As opposed to other 3D FFT implementations, CROFT uses only two threads where one thread 
is dedicated for communication so that it can be effectively overlapped with computations. Thus, depending on the number of processes 
used, CROFT  achieves performance improvement of about $51\%$ to $42\%$ as compared to FFTW3 library

\end{abstract}
\end{frontmatter}

\section{Introduction}
Fast Fourier Transform (FFT) is an  extensively used  algorithm which calculates the Discrete Fourier Transform (DFT)  of $N$ complex points. 
 Discrete Fourier Transform is the most fundamental mathematical tool  applied to time series and waveform analysis in signal processing, applied mathematics, spectral analysis, control processing etc
 \cite{Stearns1993}. With the famous divide and conquer algorithm by Cooley and Tukey \cite{Cooley1965}, the FFT algorithm reduced the time  complexity of naive implementation of  
 DFT from O($n^2$) to O$(n \log n)$ for serial computation. It also opened up an active area for parallel implementation of FFT 
 algorithms, depending on data size and machine architecture.  
 Fast Fourier Transform as a numerical tool, has been extensively used  across wide  disciplines of science and engineering. For example, its  application ranges from turbulence simulations, computational chemistry and biology, gravitational interactions, cardiac electro-physiology, acoustic, seismic and electromagnetic scattering, image processing and many other
 areas {\cite{Lee2013}, \cite{Dror2010}, \cite{Aarseth2003}, \cite{Bruno2001}, \cite{Philips1997}}.

Molecular Dynamics (MD) simulation is one of the area of computational biophysics research, which requires three dimensional FFT 
calculations to estimate energies and forces due
to long range electrostatic potential \cite{Frenkel2002} \cite{Rapaport}. 
The efficient computation of Coulomb interactions in charged particle systems is of great 
importance in this field. Particle Mesh Ewald (PME) is most common and extensively used  method to deal electrostatics calculations 
in MD simulations with periodic boundary conditions. The performance of MD simulations softwares mainly depends on efficiency of PME 
calculations and  scalability of 3D 
Fast Fourier transform (FFT). In PME method, Poisson’s equation  is solved by performing a forward FFT on charges of atoms 
distributed on three dimensional grid. Then after calculating energy in reciprocal space, reverse FFT is performed to estimate 
electrostatic force on individual atoms \cite{Frenkel2002} \cite{Rapaport}. However, the fast fourier transforms (FFTs) represent a
major scaling bottleneck for long  range electrostatic calculations when running on many cores due to 3D FFT’s high communication cost.

The most popular MD simulation codes like GROMACS \cite{Hess2008}, AMBER \cite{Case2005}  and CHARMM \cite{Brooks2009} use FFT from the FFTW3 library to perform electrostatic calculations. Most of these codes use MPI for parallelization in order to reduce the computation time on supercomputing clusters. FFT is applied on large data sets with multiple dimensions. This makes FFT calculations computationally intensive, and parallel FFT involves data distribution and collective communication. Therefore, a lot of efforts on research and development in parallelization of  FFT,  especially on 3D FFT, have been carried out for a variety of domain specific applications.
With modern day HPC environments, where large number of processors are available, scalability  and performance of 3D FFT is a major challenge.
Parallelization of FFT algorithms can be broadly categorized as distributed FFT and transpose-based algorithms \cite{Foster1997}, \cite{Grama2003}.  
In order to utilize the maximum number of processors in modern day HPC machines, transpose-based algorithms have been predominantly used in many parallel 3D FFT codes. Looking at upcoming exascale cluster architectures, the conventional
parallel 3D FFT calculations on HPC needs improvement for better performance.

 Parallel  FFT on multidimensional data can be performed as a sequence of one-dimensional transforms along each dimension. This demands data distribution, that involves a lot of communication across 
 the processors and hence, prevents the efficient usage of large number of processors for a given data size. 
The efficiently scaled  implementation of parallel 3D FFT on  new generation HPC hardware is one of the grand challenges in scientific computing. In the last two decades lots of efforts have been made to resolve this issue using different strategies. Therefore,
many parallel open source  FFT libraries exist and have been efficiently used in academia and industry as well. FFTW (Fastest Fourier Transform from West)\cite{Frigo2005} is the most widely used library in MD simulation packages, although PFFT (Parallel FFT) \cite{Pippig2013}, P3DFFT (Parallel Three-Dimensional Fast Fourier Transforms) \cite{Pekurovasky2012} and 2DECOMP\&FFT \cite{Sylvain2010} are few such libraries.  
Most of the conventional parallel 3D FFT libraries are based on 1D or slab decomposition method, which limits scaling only up to the largest dimension of multidimensional data. While using pencil or 
	2D decomposition, scalability of 3D FFT has been improved in libraries like P3DFFT and 2DECOMP\&FFT.      
All of these libraries use MPI for message passing on a distributed cluster for parallel FFT calculations. 

On the other hand, hardware reconfiguration techniques and accelerators have also been used to obtain 
performance of 3D FFT \cite{Nidhi2013},\cite{Sheng2014},\cite{Keskin2017}. Similarly, 3D parallel FFT libraries like AccFFT have been 
developed to achieve scalability and performance on both CPU and GPU architectures \cite{Gholamia2016}. 
In recent past some efforts to use combination of OpenMP and MPI have been reported to speed up 3D FFT parallel calculations on manycore/multicore hardware \cite{Nikl2014}, \cite{Takahashi2006}, \cite{Song2014}, \cite{Doi2010}.  
Although there has been a lot of research to implement 3D FFT for CPU and GPU hybrid architecture,  performance of 3D FFT implementation  on pure CPU clusters needs to be addressed.

In this paper, we present CROFT library to calculate 3D parallel FFT which is a novel way of implementation using  MPI and threaded 
programming model with pencil/2D decomposition as data decomposition strategy. In order to overlap computation and I/O operations with 
communication, we have used multithreading as opposed to conventionally used non blocking MPI calls \cite{Hollingsworth2014} \cite{Song2014}, \cite{Doi2010}. 
This library  was developed as an effort to  address the scalability bottleneck faced by conventionally used FFTW library to 
perform 3D FFT calculations in MD simulations. Where, currently the average problem size requires 3D FFT of grid dimensions 
ranging from 128 x 128 x 128 to 1024 x 1024 x 1024 \cite{Perilla2015}. Therefore we have considered these dimensions while  developing the 
CROFT  library to increase the performance and scalability on large multicore clusters. 
To exploit the multithreading capabilities of the processor, this paper presents an approach  in which  MPI along with threads is used 
such that  the overlapping of compute and memory-I/O with MPI communication is achieved by threading.
CROFT has demonstrated performance improvement of approximately $42\%$ to 
$51\%$ with varying number of processes as compared to the popularly used FFTW3 library.

\section{Background and Implementation of parallel 3D FFT}

\subsection{Multidimensional FFT}
The forward DFT of a three-dimensional complex input array $X = \{ X (0:N_{x} -1, 0: N_{y} - 1, 0:N_{z} - 1 )$ \} to a complex three-dimensional output array $Y = \{ Y (0:N_{x} -1, 0: N_{y} - 1, 0:N_{z} - 1 ) \}$ is defined as \cite{Sigrist2007},
\begin{equation}
Y(k_{x},k_{y},k_{z}) = \sum^{N_{x}-1}_{j_{x}= 0} \sum^{N_{y}-1}_{j_{y}= 0} \sum^{N_{z}-1}_{j_{z}= 0}  X(j_{x},j_{y},j_{z}) E
\end{equation}	
\begin{eqnarray}
{\textrm{where}},  & E &  =  e^{ -2 \pi \imath \left( \frac{k_x j_x}{N_x} +  \frac{k_y j_y}{N_y} +  \frac{k_z j_z}{N_z}  \right)} {\textrm{and}} \nonumber   \\	
 && 0 \leq k_x < N_x ,\nonumber \\
 && 0 \leq k_y < N_y , \nonumber \\
 && 0 \leq k_z < N_z  \nonumber
\end{eqnarray}

The corresponding backward DFT using the same definitions is defined as  \cite{Sigrist2007},
\begin{equation}
	X(j_x,j_y,j_z) = \frac{1}{N_x N_y N_z } \sum^{N_{x}-1}_{k_{x}= 0} \sum^{N_{y}-1}_{k_{y}= 0} \sum^{N_{z}-1}_{k_{z}= 0} Y(k_x,k_y,k_z ) E  
\end{equation}	
\begin{eqnarray}
{\textrm{where}}, & E & = e^{2 \pi \imath \left( \frac{k_x j_x}{N_x} + \frac{k_y j_y}{N_y} +  \frac{k_z j_z}{N_z} \right) }  {\textrm{and}}  \nonumber\\
&& 0 \leq j_x < N_x, \nonumber \\
&& 0 \leq j_y < N_y, \nonumber \\
&& 0 \leq j_z < N_z  \nonumber \\
\end{eqnarray}

For parallelization of 3D FFT, the 3D input matrix data can be decomposed and distributed amongst the processes \cite{Sigrist2007}. To perform 3D FFT  we have to take 1D FFT, along each  dimension. This can be achieved using the serial 1D FFT as the building block.

\subsection{Decomposition techniques}
There are three main  data decomposition techniques available. These are 1) slab or 1D decomposition \cite{Foster1997}, 2) pencil or 2D decomposition \cite{Ayala2012} and 3) cell or 3D decomposition \cite{Sedukhin2012}. 

\subsubsection{Slab Decomposition}

\begin{figure}
\centering	
\includegraphics[width=3in, height=2.5in]{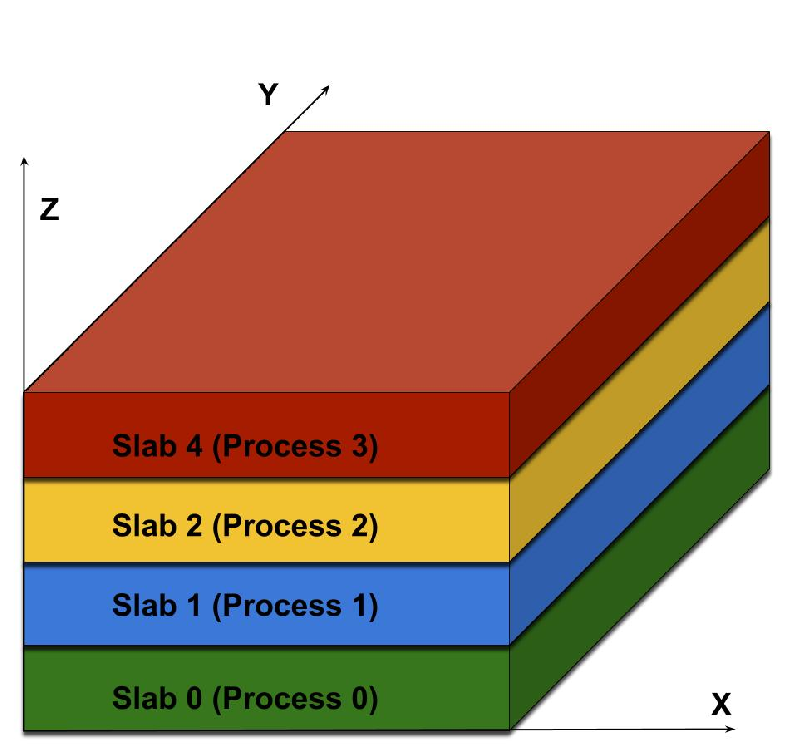}
\caption{Slab or 1D decomposition technique for parallelization.}
\label{fig:slab}
\end{figure}

In this approach 3D input matrix is decomposed along any one dimension resulting in multiple slabs which are given to different processes for further computations. For example, if input matrix has 
dimension $ N_x \times N_y \times N_z$, and any one axis, say $Z$ is chosen for decomposition, then the distributed input with every process will be,
\begin{equation}
N_x \times N_y \times \frac{N_z}{P_z} 
\end{equation}
\begin{eqnarray}
\textrm{where, } \nonumber 
 P =P_z =  \textrm{Number  of processes along $ Z$ axis.} \nonumber
\end{eqnarray}
As data along the $X$ and $Y$ is contiguous in memory, we can either take 2D FFT transform or separately take 1D FFT along both the axes locally. 
These local transforms do not involve any kind of communication between the processes.
After 2D transform along the $X$ and $Y$ dimensions,  we have to  take a global transpose of the data and then perform 1D FFT along the $Z$ axis. This global transpose is required to make the third dimension locally available on the processes and involves communication between the processes to exchange data.
The scalability of the slab decomposition is limited by the number of slabs that can be created along a single dimension of the 3D matrix. In this case, the maximum number of processes that can be used is $P_{max}=N_z$.  
Thus, this technique is not suitable when large number of processors are available.
Slab decomposition is used by many parallel 3D FFT libraries e.g. FFTW3 \cite{Frigo2005} and problem-specific applications e.g. molecular dynamics software GROMACS \cite{urlgromacs}.

\subsubsection{Pencil Decomposition}

In this approach 3D input matrix is decomposed along two dimensions which forms a shape of pencil. Number of pencils generated are equal to the number of processes to be spawned. For example, if the input 
3D matrix has dimension $N_x \times N_y \times N_z$, then take any two dimensions for decomposition, say Y and Z. The distributed input with every process will be, 

\begin{equation}
	N_x \times  \frac{N_y}{P_y} \times \frac{N_z}{P_z}
\end{equation}

\begin{eqnarray}
\textrm{where} \nonumber\\ 
P_y &=& \textrm{Number of processes along $Y$ axis,} \nonumber\\
P_z &=& \textrm{Number of processes along $ Z $ axis, and} \nonumber \\
P_{} &=& P_y \times P_z \textrm{ is the total number of available processes.} \nonumber\\ 
\end{eqnarray}
\begin{figure}
\centering
\includegraphics[width=3in, height=2.5in]{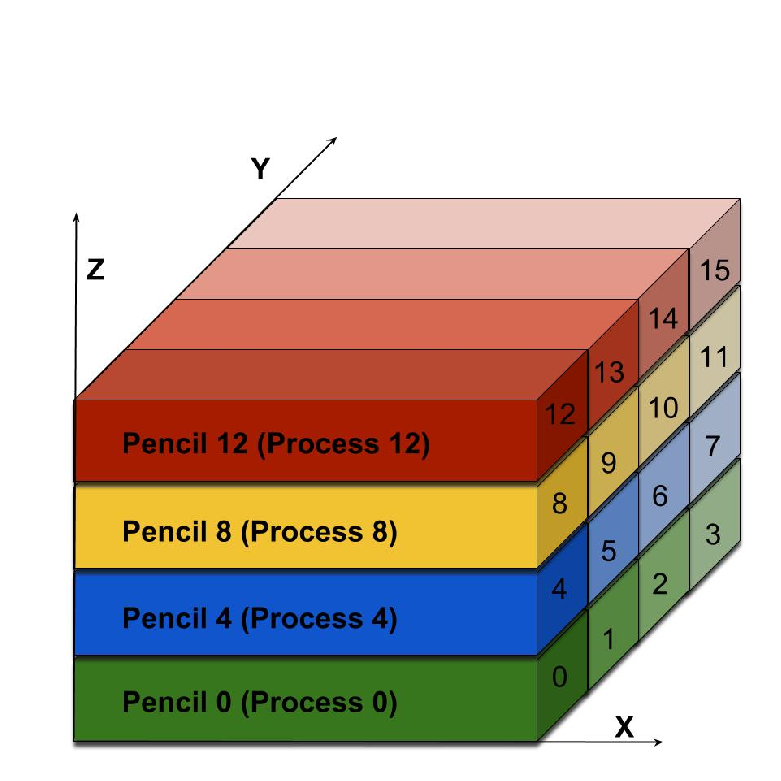}
\caption{Pencil or 2D decomposition technique for parallelization.}
\label{fig:pencil}
\end{figure}

The scaling limitation of 1D decomposition technique can be overcome by using a 2D decomposition technique. Here, maximum number of processes that can be used are $P_{max}=N_y \times  N_z$ which is greater than that of slab decomposition.

\subsubsection{Cell decomposition}

In this approach 3D input matrix is decomposed along all three  dimensions to form small cuboidal sub-matrices of data, called cells. Number of cells generated are equal to number of processes to be spawned. For example, if input 
3D matrix has dimension $N_x \times N_y \times N_z$, then distributed input with every process will be,
\begin{equation}
\frac{N_x}{P_x} \times \frac{N_y}{P_y} \times \frac{N_z}{P_z} 
\end{equation}	
\begin{eqnarray}
\textrm{where} \nonumber\\ 
P_x &=& \textrm{ Number of processes along $X$ axis,} \nonumber\\
P_y &=& \textrm{ Number of processes along $Y$ axis,} \nonumber\\
P_z &=& \textrm{ Number of processes along $Z$ axis, and} \nonumber\\
P_{} &=& P_x \times P_y \times P_z \nonumber\\
     & &\textrm{is the total number of available processes.} \nonumber 
\end{eqnarray}
	
For calculation of 3D FFT using this approach, we can use a large number of processors, but for a very large number of processes, the complexity increases as it involves more number of processes carrying out communication to exchange data.

\begin{figure}
\centering
\includegraphics[width=3in, height=2.5in]{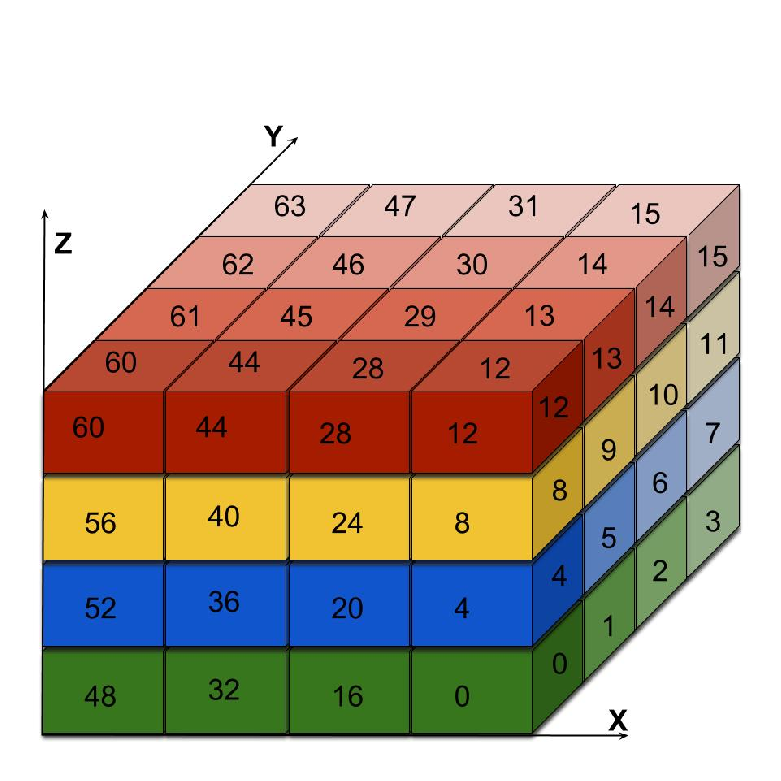}
\caption{Cell or 3D decomposition technique for parallelization.}
\label{fig:cell}
\end{figure}

\section{Related Work}

To perform parallel 3D FFT, different open source libraries  such as FFTW3, P3DFFT, and 2DECOMP$\&$FFT are available. FFTW3 uses slab decomposition, whereas 
P3DFFT and 2DECOMP$\&$FFT use pencil 
decomposition to distribute the data in parallel environment.

FFTW3 is a widely used free-software library that computes the Discrete Fourier Transform (DFT) and its various special cases \cite{Frigo2005}, \cite{urlfftw}.
FFTW3 version 3.3.8 is used for comparative study and serial 1D FFT calculations. FFTW3 uses slab decomposition and therefore its scaling is limited to $P_{max} <= N$ where, $ P_{max}$ is the maximum number of processors and $N$ is linear problem size. CROFT differs from FFTW3 in use of pencil or 2D decomposition for data distribution. It makes use of threads for overlapping compute and memory I/O with communication instead of computation optimisation as is the case with FFTW3.

P3DFFT \cite{Pekurovasky2012} uses Message Passing Interface (MPI) for interprocessor communication, and
from v.2.7.5 onwards P3DFFT provides a multi-threading option for hybrid MPI/OpenMP implementation.
In this case the MPI decomposition is the same as the non threaded version, and each MPI task now has N threads. 
This essentially implements 3D decomposition. On other hand 2DECOMP\&FFT \cite{Sylvain2010} library uses 2D or pencil decomposition for 
data distribution on distributed-memory platforms. It is claimed to be one of the scalable and efficient distributed Fast Fourier 
Transform modules that supports three-dimensional FFT. This library does not make use of multithreading.

CROFT uses multithreading for overlapping communication and computation whereas 2DECOMP\&FFT \cite{url2decomp} does not use multithreading, and multi threading used in P3DFFT is essentially to perform computation \cite{urlp3dfft}.

\section{Proposed Method}
CROFT is a parallel three-dimensional Fast Fourier Transform library implementation for distributed 
clusters. We have used a general algorithm which is based on pencil decomposition for data distribution. 
It is implemented using MPI and is based on the strategy of overlapping compute and 
communication operations. To achieve this overlap, two threads are used where one thread is dedicated for MPI communication. Three-dimensional FFT is obtained by calculating 1D FFT along all the three 
dimensions of the input data. CROFT uses 1D FFT routine from the FFTW3 library to calculate the FFT along each dimension.

\subsection{Algorithm}

\begin{figure}
\centering
\includegraphics[width=4.8in, height=2.0in]{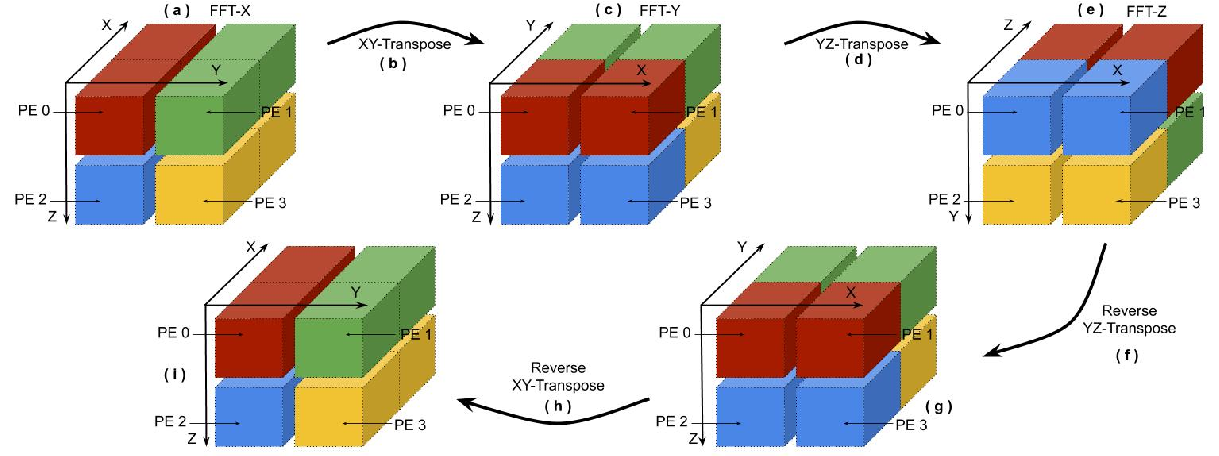}
\caption{Steps involved in computing 3D parallel FFT. Pencil image indicates 1D FFT calculations along a dimension and arrow between two images indicates transpose.}
\label{3DFFT}
\end{figure}

The algorithm requires $2^{p}$ processes which are arranged as a two dimensional
matrix. Processes in each row form row-communicator and processes in each column
form column-communicator. This results in multiple row and column communicators
as seen in figure \ref{fig:vgrid}. The algorithm requires each process to have its part of data. For
the sake of understanding, it is assumed that the data is aligned along the $X$
dimension. The data is decomposed along the $Y$ and $Z$ dimensions to form multiple
pencils, which are aligned along the X dimension as seen in figure \ref{3DFFT}(a). The number of pencils
would be equal to the number of processes, where one pencil is assigned to each process. The
steps followed by CROFT are given below.

\noindent
{\bf{Steps:}}
\begin{enumerate}
\item Compute 1-D FFT along the X dimension for all processes.
\item Pack the 1-D array data into a buffer in preparation for all-to-all communication.
\item  For all column communicators, perform all-to-all communication between all processes in a column communicator.
\item Rearrange the received data into the 1-D array with the new memory layout such that elements on the $Y$ dimension are adjacent in memory.
\item Compute 1-D FFT along the $Y$ dimension. 
\item Pack the 1-D array data into a buffer in preparation for all-to-all communication.
\item For all row communicators, perform all-to-all communication for each process in a row communicator. 
\item Rearrange the received data into the 1-D array with new memory layout such that the elements on the $Z$ dimension are adjacent in memory. 
\item Compute 1-D FFT along the $Z$ dimension.
\end{enumerate}

\subsection{Algorithm Explanation}

Initially, the data is distributed to each process as a pencil in which the data is aligned along the $X$ axis. Each process now computes 1D FFT along the $X$ axis and saves the result in 1D array. 
To compute FFT along the $Y$ axis, $XY$ transpose is performed so that data along the $Y$ axis becomes contiguous. This is achieved using step 2, 3 and 4 of the algorithm. 
 The 1-D array data is packed into a buffer, such that the buffer is filled with
data which is to be communicated, followed by all-to-all communication in the
column communicators. In both these operations, packing and MPI all-to-all communication have been overlapped. 
After the completion of communication, the data is unpacked and rearranged in the memory so
that the data along the $Y$ axis would be contiguous. 
Now the 1D FFT is computed along the $Y$
dimension and the result is saved in 1D array. The FFT along the $Z$ axis is computed by performing $YZ$
transpose. Now the data along the $Z$ axis would be contiguous as per the steps 6, 7 and 8.
This is followed by the overlap of the operations involving packing of data with MPI all-to-all communication in row communicators. 
After the completion of
communication, the data is unpacked and rearranged in memory so that it would be contiguous along the
$Z$ axis. Now, 1-D FFT is computed along the $Z$ dimension and the
result is saved in 1D array. To get the same data layout as initial, $YZ$ and $XY$
transposes are performed.
%again using overlapping of data packing with MPI all-to-all communication.

\section{Implementation and verification}
The above mentioned algorithm (Section 4.1) is implemented in CROFT library to calculate parallel 3D FFT
using pencil decomposition. The implementation is done in C using MPI along with threads
for double precision complex
data. This implementation considers the dimensions of the actual 3D matrix as $N_x$ , $N_y$
and $N_z$, where $N_x = N_y = N_z$ and is equal to $2^{n}$ for any integer $n$.
We have discussed the implementation of forward transform in this paper. The backward transform can be obtained by reversing the steps in the algorithm.

\subsection{Parallelization and optimization}
\begin{figure}
\centering
\includegraphics[width=3in, height=2.5in]{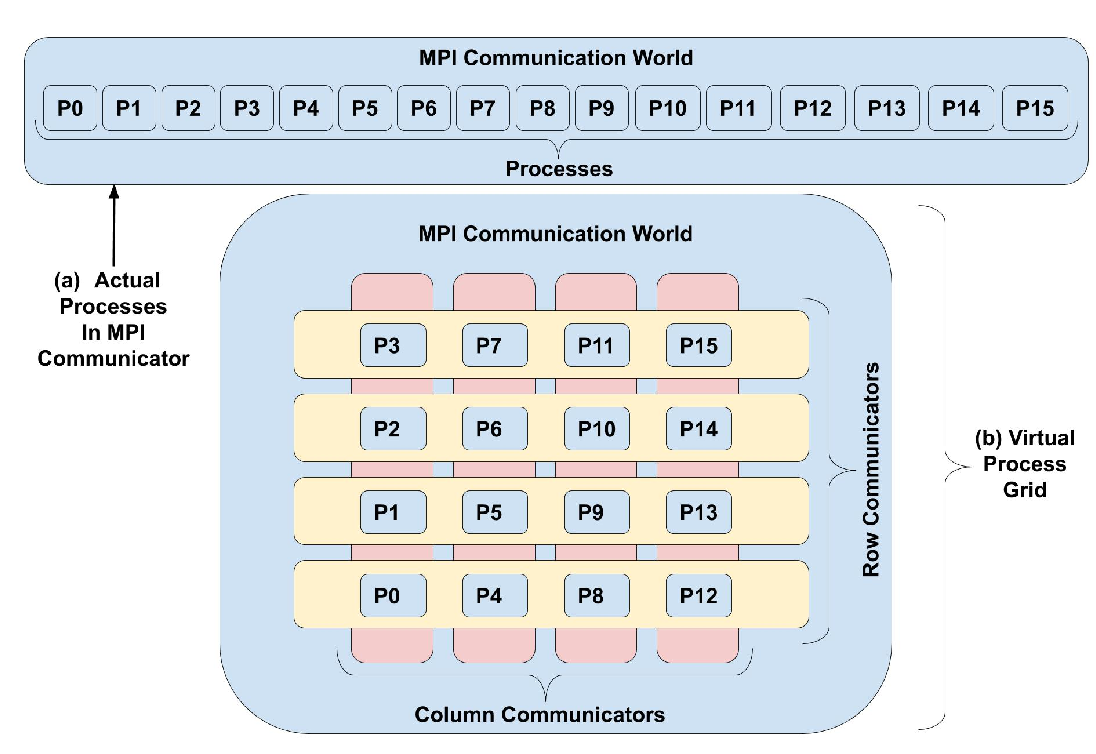}
\caption{2D virtual communication grid formed by processes in 2D or pencil decomposition}
\label{fig:vgrid}
\end{figure}

Message Passing Interface (MPI) library is used to communicate across the
processes in a distributed cluster. The total number of processes are virtually
arranged in 2D virtual communication grid as shown in figure \ref{fig:vgrid}. Here $P_y$ is
the number of processes along the $Y$ axis and $P_z$ is the number of processes
along the $Z$ axis.

Initially, as the data is contiguous along the X axis, each process first performs a 1D FFT along it.
For all the nodes to have the $Y$ dimension locally available, a global transpose is required. It then takes global transpose and performs 1D FFT along the $Y$ axis. At this stage, to swap the X and Y axis, 
all-to-all communication between processes within the same row of the virtual
communication grid takes place. It again performs global transpose and 1D
FFT along the $Z$ axis. This global transpose is required for the nodes to have
the $Z$ dimension locally available. Here, all-to-all communication between
processes within the same column of the virtual communication grid is
required to swap the data along the $Y$ and $Z$ axis.

\begin{figure}
\centering
\includegraphics[width=3in, height=1.7in]{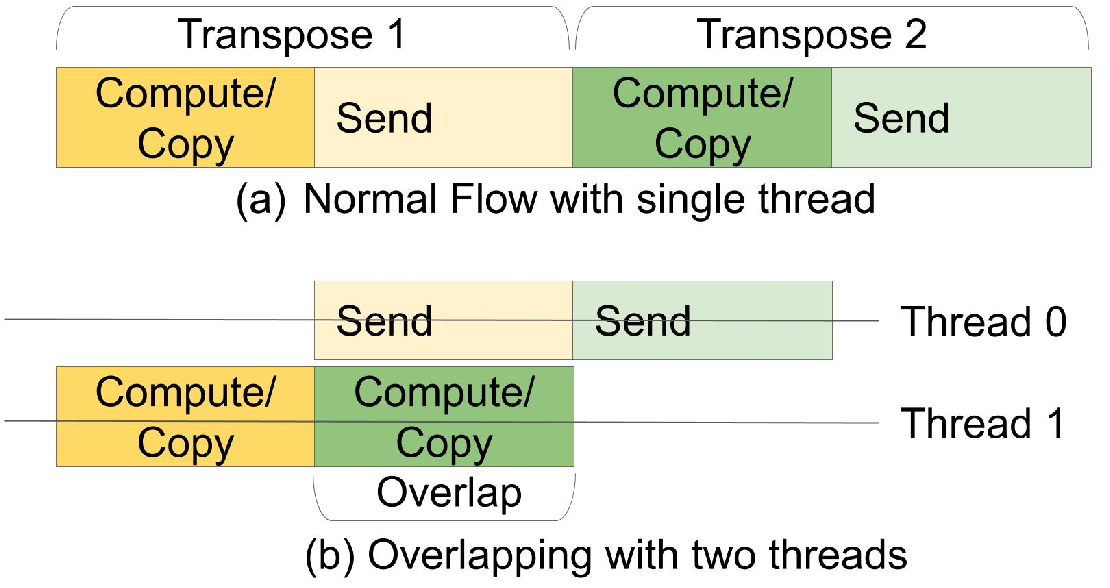}
\caption{Overlapping of data copy and communication operations.}
\label{Overlap}
\end{figure}

Threading is used for overlapping compute and memory I/O with communication operations. 
This implementation uses two threads as seen in figure \ref{Overlap}(b) with one thread
dedicated for MPI communication. While one thread is executing the application, if communication 
is demanded, it is handled by the other thread. Here, we define an iteration parameter 
$K$ to handle the trade-off between size of overlapping computation and memory I/O chunk
with the communication calls. With a small value of $K$, we can overlap a large chunk of compute 
and memory I/O operations with the communication calls, which results in less communication overhead. 
If the value of parameter $K$ gets large, we can overlap smaller chunks of compute and memory I/O
operations with the communication, which results in more number of communication calls. 
Thus we should find an optimum value of $K \ge 1$ to achieve high performance.

For optimizing the code execution, various techniques such as function
inlining and vectorization is used wherever possible. For checking the
performance with different implementation approaches, code was implemented
for various combinations of compute and communication options. These
options include:

\noindent
Option 1 : without overlap of compute and communication while using multiple
FFTW3 plans for calculating 1D FFTs.

\noindent
Option 2: without overlap of compute and communication while using single
FFTW3 plan for calculating multiple 1D FFTs.

\noindent
Option 3 : with overlap of compute and communication while using multiple
FFTW3 plans for calculating 1D FFTs.

\noindent
Option 4: with overlap of compute and communication while using single
FFTW3 plan for calculating multiple 1D FFTs.

However, as option 4 has been observed to perform better on large data and
large number of processors, “CROFT” library is implemented using option 4
with the value of $K$ fixed as 2.

\subsection{Forward transform implementation}

Forward transform is computed after data is divided in number of pencils as
seen in figure \ref{3DFFT}(a). Every process will get its own chunk of 3D pencil data
($N_x$, $N_y/P_y$, $N_z/P_z$ according to pencil decomposition) as a 1D input array.
Each process executes the above given algorithm with the help of two OpenMP threads which are
spawned after the first step. The thread with thread Id 0 is used for MPI
communication and the thread with thread Id 1 is used for packing the data
into a buffer and performing 1D FFT as seen in figure \ref{Overlap}(b). Two threads run
simultaneously and achieve overlapping of computation and memory I/O with
all-to-all communication, i.e step 2 and step 3 of the algorithm are successfully
overlapped. After the communication the data is unpacked and FFT along the
$Y$ direction is computed. Once again packing of data and all-to-all
communication steps are overlapped in preparation for the FFT along the last
dimension. Then the data received after communication is unpacked and FFT
along the $Z$ direction is computed. To get the same data layout as initial, 
we again perform $YZ$ and $XY$ transpose respectively and use overlapping of data packing with
MPI all-to-all communication.

\subsection{Verification of code}

3D parallel FFT using pencil decomposition requires MPI all-to-all
communication and many data copy operations within local memory for
rearranging the data. Therefore, it is necessary to verify the result after every
step. For verification purpose, we first implemented routines to
print the results and verified the generated output data with the desired output for the given input.
Secondly, we have taken backward FFT to get back the original input. Since,
we did not perform any manipulations while using normalization factor,
output of backward transform is same as the input applied. Finally, to check the correctness, we tested the results obtained by CROFT
library against the results from FFTW3 library for double precision complex
input. The output was found to be exactly the same.

\section{Discussion and Results}	

\subsection{Benchmarking details}
\subsubsection{Experimental Setup}
CROFT was benchmarked against the 3D FFT API from
FFTW3 v3.3.8 on the Param
Bioblaze cluster and Sangam Lab cluster which are internal clusters
in C-DAC. The Param Bioblaze cluster is a blade based cluster with two chassis
which are interconnected with an external 56 Gbps Mellanox FDR IB switch.
Each chassis contains 16 dual socket blade servers connected through an internal
FDR IB switch. Each blade server has two 8 core Intel sandy bridge processors
and 64 GB RAM. So, the number of cores in the Param Bioblaze cluster accumulates to a total of 512 compute cores.
CROFT library and other applications are compiled with Intel MPI version 14.0.2 with -O2 optimisation flag enabled.
CROFT uses two threads per process which is fixed. One thread performs only communication and the other thread performs all other tasks.

\subsubsection{Input data}
The input data used for the benchmarking purpose was the 3D matrix of double precision complex numbers.
For performing the benchmark runs, we spawned one process per core for a smaller 3D matrix of size $128 \times 128 \times 128$.
For a larger 3D matrix of size $1024 \times 1024 \times 1024$ less number of processes per node were used keeping few cores idle due to memory constraint.

\subsubsection{Time measurements}

The timing information is collected for benchmarking purpose using MPI\_Wtime API. The starting 
timestamp is collected just before calling the 3D FFT API of CROFT and FFTW3 library functions. The 
processes are synchronized at a global barrier to avoid distortion of the time before collecting the 
initial timestamps. Another timestamp is collected just after the 3D FFT API execution is completed. 
The difference between the two timestamps is considered as the execution time required for the process 
to perform 3D FFT. We then get the minimum and maximum execution time taken by the processes into 
process 0 using a global reduction with MPI\_MAX and MPI\_MIN options. The time obtained from 
MPI$\_$MAX reduction is considered as the wall time required by the 3D FFT library function. 
To get the final wall time, multiple runs of application are done and the avarge timings are selected.

\subsection{Results}
Benchmarking runs were performed on parallel 3D API of FFTW3 and all the
implemented options as discussed in Section 5.1.

\begin{table}
\begin{center}
\begin{tabular}{ | c | c | c | c | c | c|}
\hline 
\multirow{2}{*}{Number of} & \multicolumn{5}{c|}{Time in (sec)} \\
cores	&  \multicolumn{5}{c|}{}  \\
\cline{2-6}	
	& FFTW3  & opt1  & opt2 & opt3 & opt4 \\
\hline 
4 & 0.053 & 0.163  & 0.060 & 0.166 & 0.045 \\
\hline

8 & 0.029 & 0.089 & 0.037 & 0.097 & 0.036 \\
\hline

16 & 0.020 & 0.055 & 0.029 & 0.060 & 0.032 \\
\hline

32 & 0.019 & 0.028 & 0.014 & 0.031 & 0.017 \\
\hline

64 & 0.494 & 0.038 & 0.031 & 0.039 & 0.032 \\
\hline

128 & 1.911 & 0.037 & 0.031 & 0.039 & 0.032 \\
\hline

256 & 5.473 & 0.073 & 0.069 & 0.074 & 0.076 \\
\hline

512 & 26.149 & 0.183 & 0.178 & 0.126 & 0.123 \\
\hline
\end{tabular}
\end{center}
\caption{{Timings (in sec) on Param Bioblaze cluster for benchmarking with 3D matrix of size $128 \times 128 \times 128$ and nodes fill up allocation for FFTW3 and multiple options of CROFT. Opt 1: Without overlap multiple plans; Opt 2: Without overlap single plan; Opt 3: With overlap multiple plans; Opt 4: With overlap single plan}}
\label{size128}
\end{table}

\begin{table}
\begin{center}
\begin{tabular}{ | c | c | c | c | c | c| c |}
\hline 
Layout  & Number of  & \multicolumn{5}{c|}{Time in (sec)}\\
(Nodes $\times$ &   cores  & \multicolumn{5}{c|}{}      \\
\cline{3-7}	
ppn)        &           & FFTW3      & opt1        & opt2    & opt3    & opt4    \\

\hline 
2 $\times$ 2 & 4 & 0.057 & 0.156 & 0.043 & 0.166 & 0.041 \\
\hline

4 $\times$ 2 & 8 & 0.032 & 0.079 & 0.025 & 0.082 & 0.021 \\
\hline

4 $\times$ 4 & 16 & 0.021 & 0.044 & 0.017 & 0.043 & 0.016 \\
\hline

4 $\times$ 8 & 32 & 0.019 & 0.027 & 0.013 & 0.039 & 0.029 \\
\hline

8 $\times$ 8 & 64 & 0.401 & 0.037 & 0.023 & 0.065 & 0.038 \\
\hline

8 $\times$ 16 & 128 & 1.911 & 0.037 & 0.031 & 0.039 & 0.032 \\
\hline

16 $\times$ 16 & 256 & 5.473 & 0.073 & 0.069 & 0.074 & 0.076 \\
\hline

32 $\times$ 16 & 512 & 26.149 & 0.183 & 0.178 & 0.126 & 0.123 \\
\hline
\end{tabular}
\end{center}
\caption{Timings (in sec) on Param Bioblaze cluster with processes layout.}
\label{size128_2}
\end{table}

Table \ref{size128} shows the benchmarking timings on Param Bioblaze cluster with fill up
allocation, i.e. all cores on a node are used before spawning to the next node, and table \ref{size128_2} shows the timings obtained by custom layout of processes. For the 3D matrix of size $128 \times 128 \times 128$, FFTW3 code can use up to 128 cores due to the use of slab decomposition and it is evident from the time taken by FFTW3 code for more than 128 cores. 
All other implemented options take relatively less time for execution and scale upto all the 
available 512 cores. With a different layout, the timing improves to some extent, as seen in table \ref{size128_2}.

\begin{table}
\begin{center}
\begin{tabular}{ | c | c | c | c | c | c|}
\hline 
Number of &\multicolumn{5}{c|}{Time (sec)} \\
\cline{2-6}	
cores	& FFTW3  & opt1  & opt2 & opt3 & opt4 \\
%cores  \\
\hline 
4 & 101.8044 & 67.1240  & 58.7549 & 61.4341 & 51.9448 \\
\hline

8 & 51.8731 & 38.0646 & 33.8035 & 36.7066 & 29.7946 \\
\hline

16 & 26.9177 & 25.0881 & 23.2374 & 25.8118 & 24.0778 \\
\hline

32 & 13.3403 & 13.8911 & 12.8060 & 13.0762 & 12.1718 \\
\hline

64 & 8.5792 & 8.1453 & 7.6859 & 7.2702  & 6.6610 \\
\hline

128 & 5.0288 & 4.1973 & 3.9772 & 3.5217 & 3.3376 \\
\hline

256 & 4.7209 & 2.4439 & 2.2346 & 2.1265 & 1.9747 \\
\hline

512 & 18.8849 & 1.3945 & 1.3237 & 1.4165 & 1.2722 \\

\hline
\end{tabular}
\end{center}
\caption{{Timings (in sec) on Param Bioblaze cluster for benchmarking with 3D matrix of size $1024 \times 1024 \times 1024$ with nodes fill up allocation for FFTW3 and multiple options of CROFT. Opt 1: Without overlap multiple plans; Opt 2: Without overlap single plan; Opt 3: With overlap multiple plans; Opt 4: With overlap single plan}}
\label{size1024}
\end{table}

\begin{figure}
\centering
\includegraphics[width=5.0in, height=3.0in]{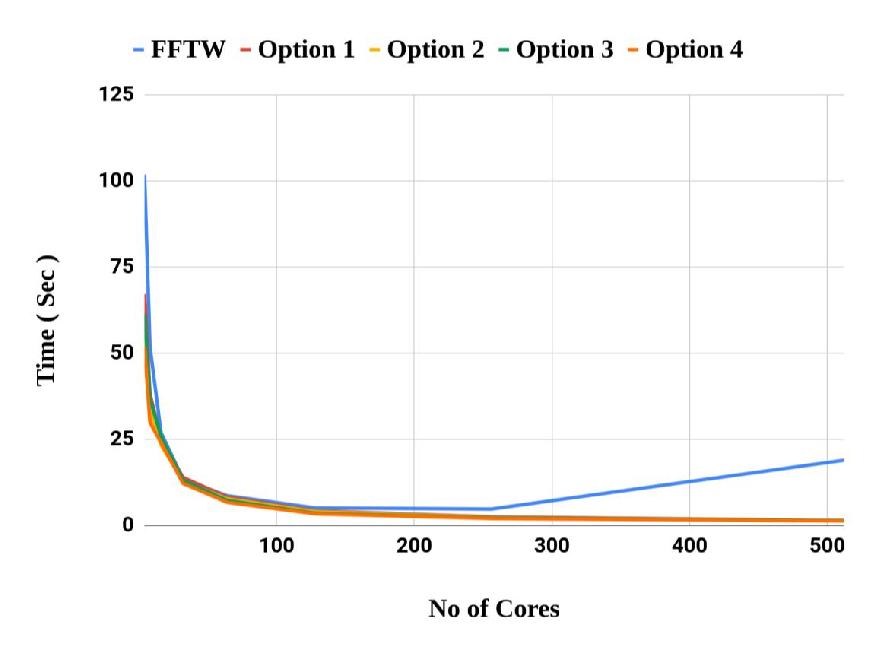}
\caption{Comparative timing chart for data size $1024 \times 1024 \times 1024$ on Param Bioblaze cluster}
\label{time_Bioblaze}
\end{figure}

For larger 3D matrix of size $1024 \times 1024 \times 1024$, CROFT implementation with overlapping of 
compute and communication along with the use of single FFTW3 plan for computing 1D FFT performs better 
than FFTW3 and all other implemented options as seen in table \ref{size1024} and figure \ref{time_Bioblaze}.

\begin{figure}
\centering
\includegraphics[width=5.0in, height=3.0in]{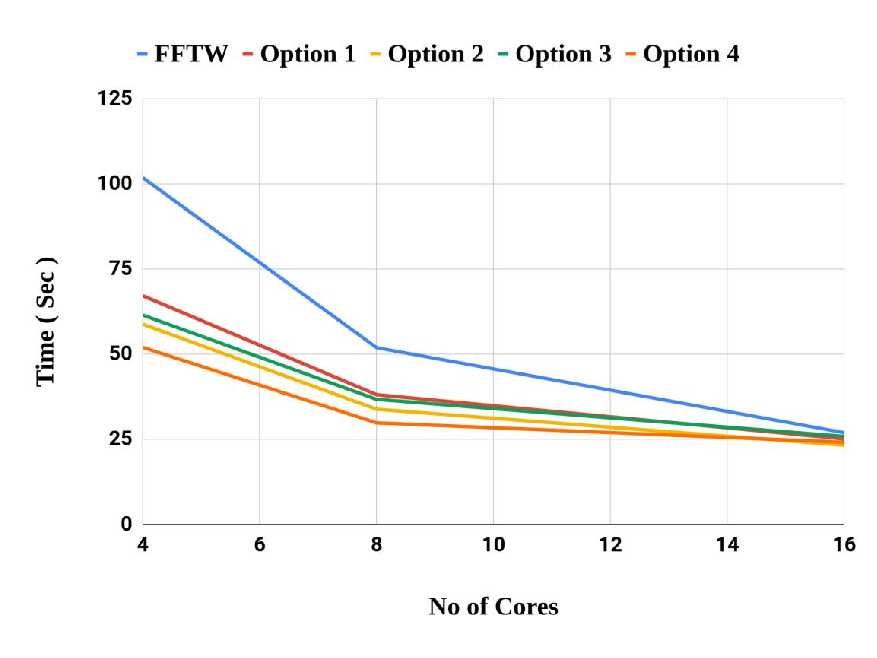}
\caption{Timings for no of cores between 4 to 16 in figure ~\ref{time_Bioblaze} }
\label{Time4-16}
\end{figure}

\begin{figure}
\centering
\includegraphics[width=5.0in, height=3.0in]{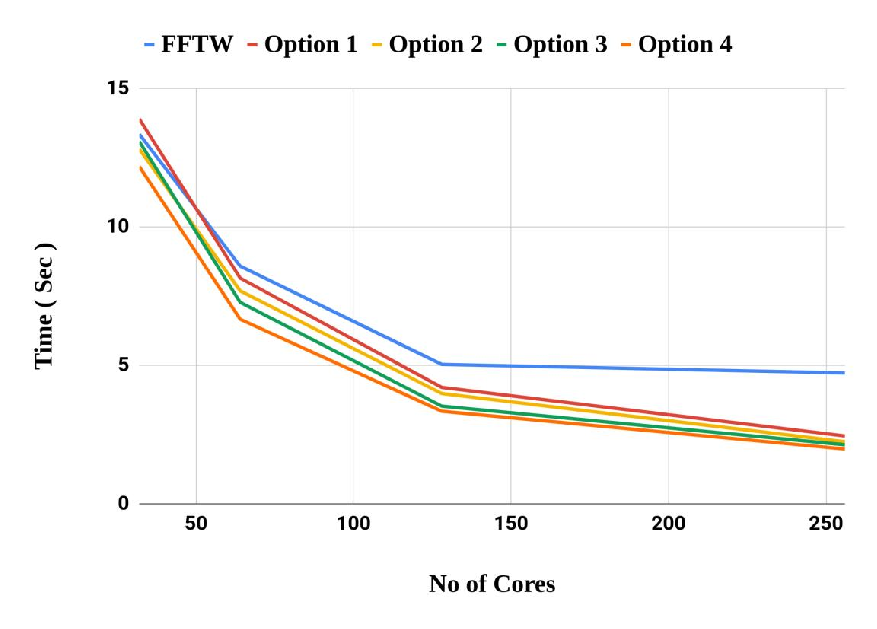}
\caption{Timings for no of cores between 32 to 256 in figure ~\ref{time_Bioblaze} }
\label{Time32-256}
\end{figure}

At smaller number of cores, the difference in execution time is more (Figure \ref{Time4-16}).
The CROFT implementation (option 4) is faster than FFTW3 by
approximately $51 \%$ when the number of cores are less (4 processes) (Figure \ref{Time4-16}) and approximately
$42 \%$ when FFTW3 is having lowest time (at 256 processes) as seen in figure \ref{Time32-256}.
\begin{figure}
\centering
\includegraphics[width=5.0in, height=3.0in]{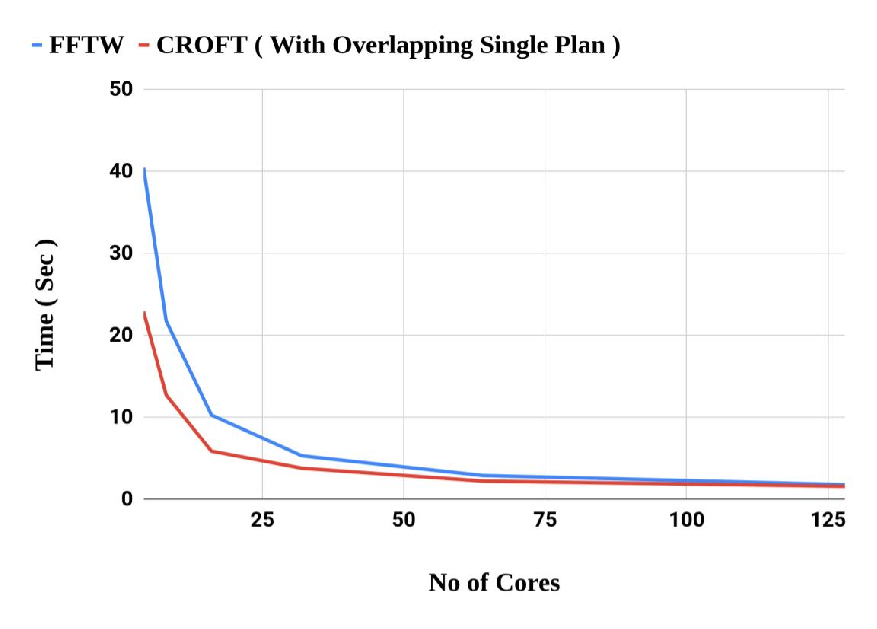}
\caption{Comparative timing chart for data size $1024 \times 1024 \times 1024$ on Sangam Lab cluster}
\label{time_Sangam}
\end{figure}
Similar results in terms of timings have been observed in Sangam Lab cluster where, CROFT (option 4) is faster than FFTW3 as seen in figure \ref{time_Sangam}.

\begin{figure}
\centering
\includegraphics[width=5.0in, height=3.0in]{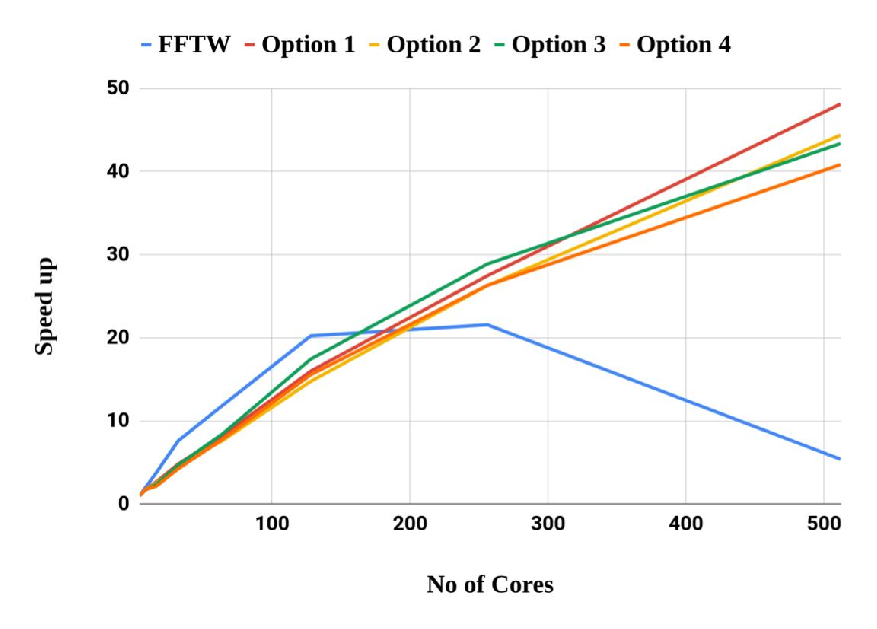}
\caption{Speedup graph for data size 1024 x 1024 x 1024 on Param Bioblaze cluster}
\label{SpeedUp}
\end{figure}

From the scalability chart as shown in figure \ref{SpeedUp}, we can see that all the
implemented options of CROFT are scalable upto all the available 512 compute cores in Param Bioblaze 
cluster whereas performance of FFTW3 drops after 128 cores.

\subsection{Profiling details}

\begin{figure}
\centering
\includegraphics[width=4.0in, height=3.0in]{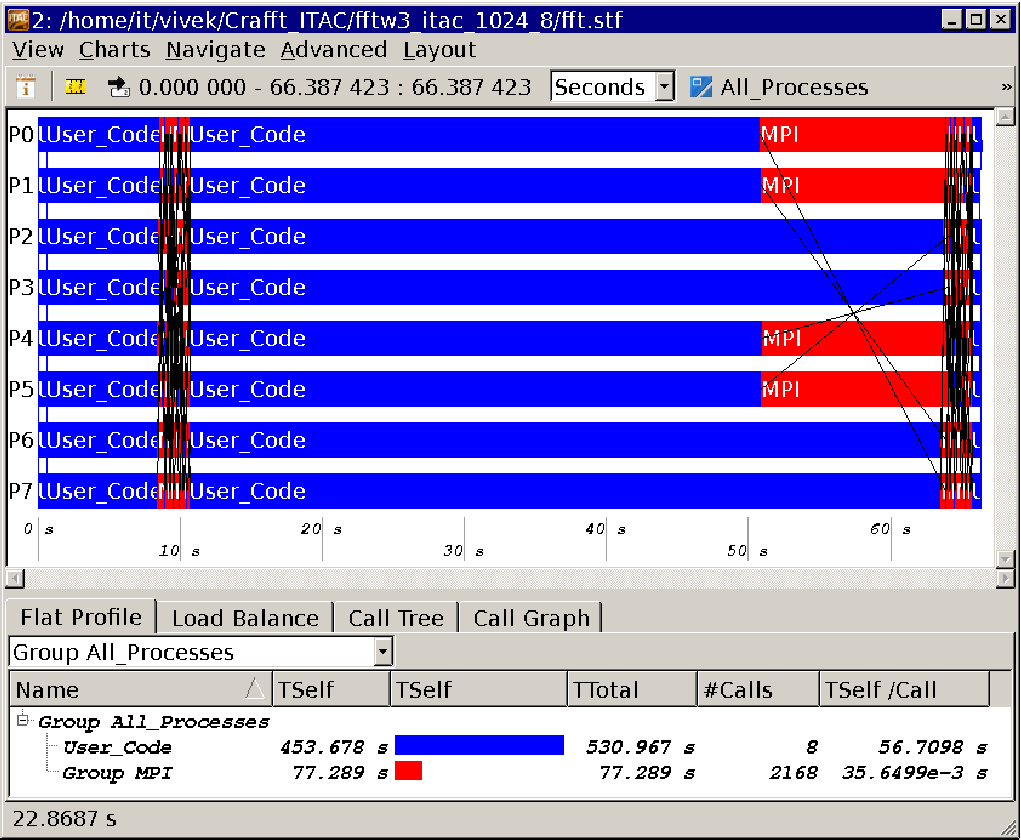} 
\caption{Event profile of “fftw\_mpi\_plan\_dft\_3d” API for 8 processes and input size of $1024 \times 1024 \times 1024$ using Intel trace collector and analyzer(ITAC)}
\label{fftw3prof_time}
\end{figure}

\begin{figure}
\centering
\includegraphics[width=4.0in, height=3.0in]{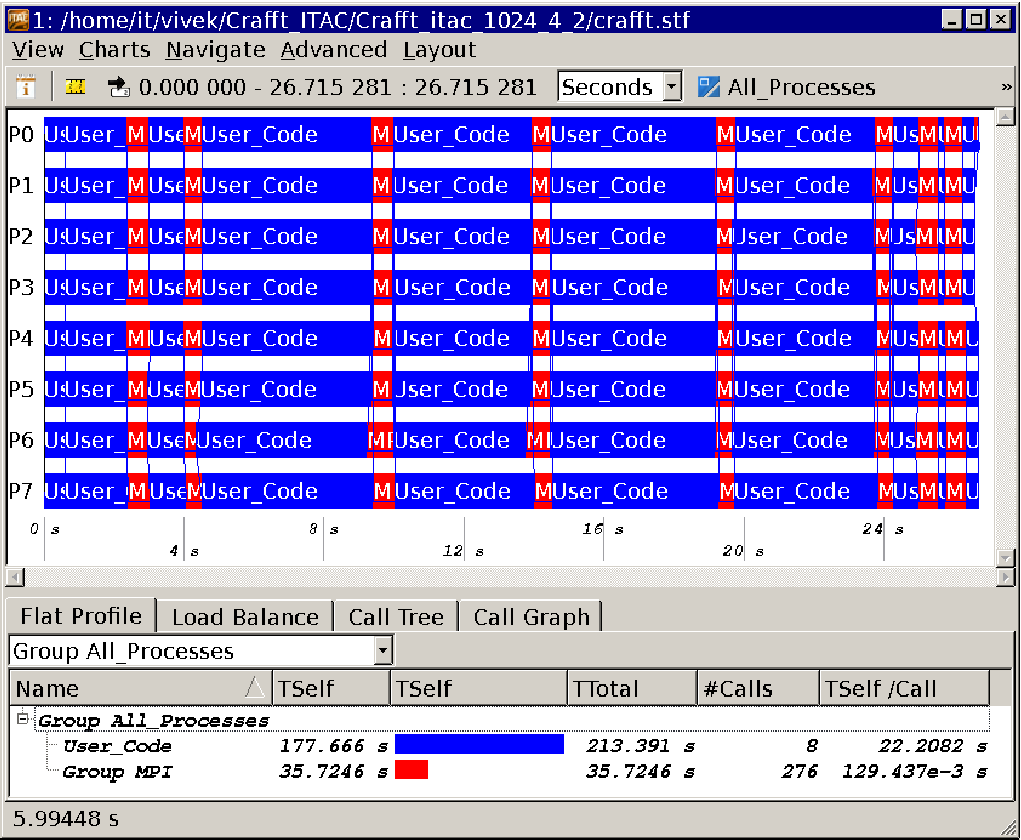} 
\caption{Event profile of “croft\_parallel3d” API for 8 processes and input size of $1024 \times 1024 \times 1024$ using Intel trace collector and analyzer(ITAC) }
\label{crafftprof_time}
\end{figure}

To get an insight on the difference between FFTW3 parallel 3D routine’s execution
and CROFT execution, profiling of both the applications have been performed on 8
processes with input matrix of size $1024 \times 1024 \times 1024$. The profiling result is
shown in figure \ref{fftw3prof_time} and \ref{crafftprof_time}. From the profiling data, 
it is clear that CROFT takes less time as compared to FFTW3. 
FFTW3 takes 453.678 seconds for the user code to execute whereas, CROFT takes only 177.666 seconds. Similarly, MPI calls take 77.289 seconds in case of FFTW3 and 35.725 seconds 
in case of CROFT as seen in figure \ref{fftw3prof_time} and \ref{crafftprof_time} respectively.

\begin{figure}
\centering
\includegraphics[width=4.0in, height=3.0in]{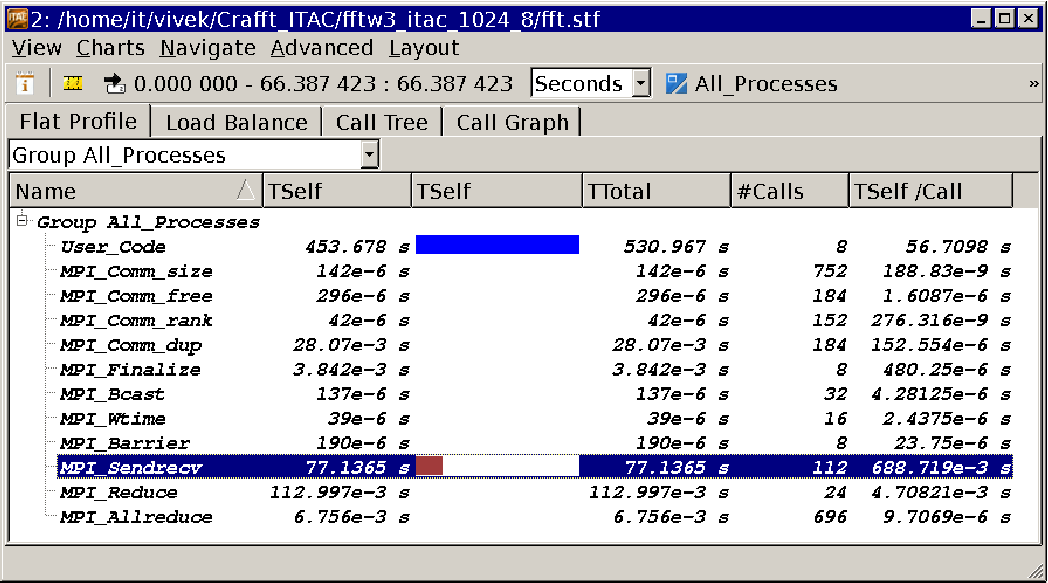}
\caption{Function profile of “fftw\_mpi\_plan\_dft\_3d” API for 8 processes and input size of $1024 \times 1024 \times 1024$ using Intel trace collector and analyzer(ITAC) }
\label{fftw3mpi_call}
\end{figure}

\begin{figure}
\centering
\includegraphics[width=4.0in, height=3.0in]{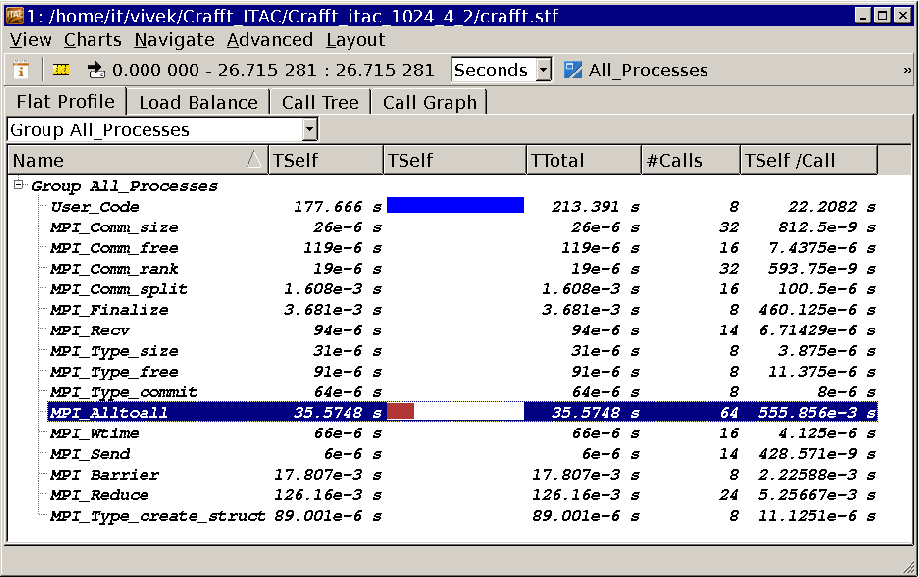}
\caption{Function profile of “croft\_parallel3d” API for 8 processes and input size of 1024 x 1024 x 1024 using Intel trace collector and analyzer(ITAC) }
\label{cafftmpi_call}
\end{figure}

Scalability of the code may be explained by analyzing the MPI communication. The code is well load 
balanced with very less waiting time for MPI calls as seen in figure \ref{crafftprof_time}. Number of 
MPI calls required by FFTW3 for communication are 864 which are much more than 124 MPI calls required by CROFT.
Apart from other MPI routines, there are 
112 MPI\_Sendrecv calls in FFTW3 routine which takes 77.136 seconds (figure \ref{fftw3mpi_call}) as compared to 64 MPI\_Alltoall calls 
used in CROFT which takes 35.574 seconds (figure \ref{cafftmpi_call}) to execute. 
The reduction in number of MPI communication calls in CROFT, indicated that the code is more scalable.

\section{Conclusion}
For the smaller datasets, FFTW3 is faster when number of cores used are less than 32, but CROFT code 
implemented with option 4 (with overlap of compute and communication while using single FFTW3 plan for 
calculating 1D FFTs) performed better when number of cores are more than 32. For larger dataset, 
CROFT implementation option 4 is the best implementation with the performance improvement between 
$42\%$ - $51\%$ as seen in table \ref{size1024}. It also scales to more  cores than FFTW3 due to 
pencil decomposition and further reducing the execution time. The shorter execution time of the CROFT may be attributed to the use of dedicated thread for communication. This implementation can be considered as a significant engineering improvement over existing FFT codes.
It can be used as one of the options for implementing exascale applications  such as MD simulations which require 3D parallel FFT.

\section{Future work}
CROFT library is a pure CPU implementation and can be extended to add support for the
accelerators like GPUs. Currently, CROFT uses 1D FFT from FFTW3 package, but
native implementation of 1D FFT can be done as a replacement to 1D FFT from
FFTW3 package, eliminating the dependency on FFTW3. Moreover, there is scope
for further memory optimization which can be looked at.

\section{Acknowledgements}
CROFT library is developed under the project National Supercomputing Mission (NSM), Government of India. 
The authors would like to acknowledge the use of Bioinformatics Resources and Applications Facility (BRAF) at the Centre for Development of Advanced Computing (C-DAC), Pune and NSM Sangam Lab cluster at the Centre 
for Development of Advanced Computing (C-DAC), Pune for testing and benchmarking of CROFT.
Authors would also like to thank Ms. Shruti Koulgi, Ms. Sunitha Manjari and Dr. Uddhavesh Sonavane for their encouragement and support.

\end{document}